\documentclass{jfm}
\usepackage{color}
\usepackage{amstext}
\usepackage{amssymb}
\usepackage{graphicx}
\usepackage{amsmath}

\newcommand\vI{\boldsymbol{I}}

\newcommand\vf{\boldsymbol{f}}

\newcommand\vl{\boldsymbol{l}}

\newcommand\vx{\boldsymbol{x}}

\newcommand\vg{\boldsymbol{g}}

\begin{document}

{\title[On self-propulsion of $N$-sphere micro-robot] {On self-propulsion of $N$-sphere micro-robot}}

\author[V. A. Vladimirov]
{V.\ns A.\ns V\ls l\ls a\ls d\ls i\ls m\ls i\ls r\ls o\ls v}

\affiliation{Dept of Mathematics, University of York, Heslington, York, YO10 5DD, UK}


\setcounter{page}{1}\maketitle \thispagestyle{empty}

\begin{abstract}

The aim of this paper is to describe the self-propulsion of a micro-robot (or micro-swimmer) consisting of
$N$ spheres moving along a fixed line. The spheres are linked to each other by arms with the lengths changing
periodically. For the derivation, we use the asymptotic procedure containing the two-timing method and a
distinguished limit. We show that in the main approximation, the self-propulsion velocity appears as a linear
combination of velocities of all possible triplets of spheres. Velocities and efficiencies of three-, four-,
and five-swimmers are calculated.

\emph{The paper is devoted to H.K.Moffatt, who initiated the author's
interests in low-Reynolds-number fluid dynamics.}

\end{abstract}

\section{Introduction and formulation of problem\label{sect01}}

\subsection{Introduction}
The studies of micro-robots represent a flourishing modern research topic, which creates a fundamental base
for modern applications in medicine and technology, see
\cite{Purcell, Koelher, NG+, Dreyfus, Yeomans1, Paunov, Lefebvre,  Gilbert, Golestanian, Golestanian1, Yeomans,
Pietro, Lauga}. At the same time, the simplicity of both time-dependence and geometry   represents the major
advantage in the studies of micro-robots (in contrast with extreme complexity of self-swimming
microorganisms, \emph{e.g.} \cite{PedKes, VladPedl, Pedley, Polin}); which makes it possible to describe the
motions of micro-robots in greater depth.

In this paper, we generalize the theory of a three-sphere micro-robot of \cite{NG+, Golestanian} to an
$N$-sphere micro-robot. We employ two-timing method and distinguished limit arguments, which lead to a simple
and rigorous analytical procedure. Our calculation of the self-propulsion velocity of an N-sphere robot shows
that it represents (in the main approximation) a linear combination of velocities due to all possible
triplets of spheres. The velocities and Lighthill's swimming efficiencies of three-,  four-, and five-sphere
robots are calculated as examples. This paper represents further studies of a problem introduced in
\cite{VladimirovX3}.

\subsection{Formulation of problem\label{sect01}}
\begin{figure}
\centering\includegraphics[scale=0.7]{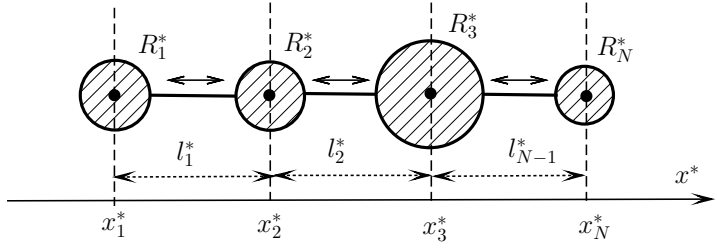}
\caption{$N=4$ spheres, linked by arms of periodically changing lengths.}
\label{N-robot}
\end{figure}

We consider a micro-robot consisting of $N$ rigid spheres of radii $R_i^*$, $i=1,2,\dots N$ with their
centers at the points $x_i^*(t^*)$ of $x^*$-axis ($x_{i+1}^*>x_{i}^*$), ${t}^*$ is time, asterisks mark
dimensional variables and parameters. The spheres are connected by $N-1$ arms/rods, such that the distances
between the centres of neighbouring spheres are $l_{\alpha}^*=|x_{\alpha+1}^*-x_{\alpha}^*|$,
$\alpha=1,2,\dots N-1$, see Fig.\ref{N-robot} (in this paper latin subscripts change from $0$ to $N$, while
greek subscripts change from $0$ to $N-1$).
We accept Stokes's approximation where masses of spheres and arms are zero; the arms are so thin (much
thinner than any $R_i^*$) that their interactions with a fluid are negligible. The equations of motion can be
written as
\begin{eqnarray}
&&f_i^*+F_i^*=0\label{exact-1}\\
&&F_i^*=-\kappa_i^*\{\dot x_i^*-
\sum_{k\neq i} 3 R_k^* \dot x_k^*/(2x_{ik}^*)\},\quad x_{ik}^*\equiv |x_i^*-x_k^*|\label{exact-1A}\\
&&x^*_{\alpha+1}-x_\alpha^*=l_\alpha^*\label{exact-1AA}\\
 &&\sum_{i=1}^Nf_i^*=0\label{exact-1B}
\end{eqnarray}
where $\kappa_i^*\equiv 6\pi\eta R_i^*$, $\eta$ is viscosity, dots above the functions stand for $d/dt^*$,
and summation convention over repeating subscripts is \emph{not in use}. The forces $f_i^*$ are exerted by
the arms on the $i$-th sphere, while $F_i^*$ represent viscous friction. In order to derive (\ref{exact-1A})
we use a classical expression for Stokes's friction force as well as the explicit expression for fluid
velocity for a moving along $x^*$-axis  sphere. The $x^*$-component $u^*$ of this velocity  at  distance
$r^*$ along $x^*$-axis is
$$
u^*\simeq 3R^*U^*/(2r^*)
$$
where $R^*$ and $U^*$ are the radius and velocity of a sphere, see \cite{Lamb,Landau,Moffatt}. Equality
(\ref{exact-1B}) follows from the fact that the external forces exerted on each arm are negligible. The
geometrical configuration of a micro-robot is determined by
\emph{given} functions
\begin{eqnarray}\label{constraint}
&&l_{\alpha}^*=L_{\alpha}^*+\widetilde{l}_{\alpha}^*(\tau)
\end{eqnarray}
where $L_{\alpha}^*$ are mean values and $\widetilde{l}_{\alpha}^*(\tau)$ are oscillations, which represent
$2\pi$-periodic functions of a dimensionless variable $\tau\equiv\omega^* t^*$; $\omega^*$ is a constant.
Since all $l_{\alpha}^*$ are given, then conditions (\ref{exact-1AA}) can be considered as geometrical
\emph{constraints}. Equalities (\ref{exact-1})-(\ref{constraint})
represent the system of $2N$ equations for $2N$ unknown functions:
$$
\vx^*(t^*)\equiv(x_1^*(t^*), x_2^*(t^*),\dots, x_N^*(t^*) ),
\quad \vf^*(t^*)\equiv(f_1^*(t^*), f_2^*(t^*),\dots, f_N^*(t^*) )
$$
These equations contain three characteristic lengths: distance $L^*$ between the neighbouring spheres, radius
$R^*$ of spheres, and  amplitude $\lambda^*$ of arm oscillations; the characteristic time-scale is $T^*$:
$$
R^*\equiv\sum_{i=1}^N R_i^*/N,\quad L^*\equiv\sum_{\alpha=1}^N L_\alpha^*/N,\quad
\lambda^*\equiv\sum_{\alpha=1}^N \max\widetilde{l}_\alpha^*/N,\quad T^*\equiv
1/\omega^*\nonumber
$$
The dimension of $\kappa^*$ can be eliminated from the equations by  division of (\ref{exact-1}) by
$\kappa^*$, hence it does not play any role in scaling. We choose dimensionless variables and small
parameters as:
\begin{eqnarray}
&& \vx^*=L^*\vx,\quad R_i^*=R^* R_i,\quad
 \widetilde{l}_{\alpha}^{*}=\lambda^*\widetilde{l}_{\alpha},\quad  t^*=T^*t,\quad f_i^*=6\pi\eta R^*L^*f_i/T^*;
 \nonumber\label{scales}\\
&&\varepsilon\equiv \lambda^*/L^*\ll 1,\quad \delta\equiv 3R^*/(2L^*)\ll 1\nonumber
\end{eqnarray}
Then the dimensionless form of (\ref{exact-1})-(\ref{constraint}) is
\begin{eqnarray}
&&R_i x_{it}-\delta\sum_{k\neq i} R_{ik}  x_{kt}/l_{ik}=f_i,
\quad l_{ik}=L_{ik}+\varepsilon\widetilde{l}_{ik}\label{exact-1-dless}\\
&&x_{\alpha+1}-x_\alpha=l_\alpha\label{exact-1AAA}\\
&&\sum_{i}f_i=\vf\cdot\vI=0,\quad \vI\equiv(1,1,\dots,1)\label{exact-1aa}
\end{eqnarray}
where $R_{ik}\equiv R_i R_k$, subscript $t$ stands for $d/dt$,
\begin{eqnarray}\label{l-ik}
l_{ik}\equiv\sum_{n=i}^{k-1} l_{n},\quad \text{for}\quad k\geq i+1,\quad\text{with}\quad
l_{ki}=l_{ik}\quad\text{otherwise},
\end{eqnarray}
and similar definitions for $L_{ik}$ and $\widetilde{l}_{ik}$ (for example, $L_{13}=L_{1}+L_{2}$
\emph{etc.}). Eqn.(\ref{exact-1-dless}) can be rewritten as $(N\times N)$-matrix form:
\begin{eqnarray}\label{exact-1-dless-matrix}
&&\mathbb{A}\vx_t=\vf\quad \text{or}\quad \sum_{k=1}^N A_{ik} x_{kt}=f_i\\
&& \mathbb{A}=A_{ik}=
\begin{cases}
R_i  & \text{ for }\ i=k,\\[-1mm]
-\delta R_{ik}/l_{ik} & \text{ for }\ i\neq k
\end{cases}\label{matrixA}
\end{eqnarray}

\section{Two-timing method and asymptotic procedure \label{sect04}}

\subsection{Functions and notations}

The following \emph{dimensionless} notations and definitions are in use:

\noindent
(i) $s$ and $\tau$ denote slow time and fast time;  subscripts $\tau$ and $s$ stand for  related partial
derivatives.

\noindent
(ii) A dimensionless function, say $G=G(s,\tau)$, belongs to class $\cal{I}$ if $G={O}(1)$ and all  partial
$s$-, and $\tau$-derivatives of $G$ (required for our consideration) are also ${O}(1)$. In this paper all
functions belong to   class $\cal{I}$, while all small parameters appear as explicit multipliers.

\noindent
(iii) We consider only \emph{periodic in $\tau$ functions} $
\{G\in  \mathcal{P}:\quad G(s, \tau)=G(s,\tau+2\pi)\},
$ where $s$-dependence is not specified. Hence all considered below functions belong to $\cal{P}\bigcap
\cal{I}$.

\noindent
(iv) For  arbitrary $G\in \cal{P}$ the \emph{averaging operation} is:
\begin{eqnarray}
\langle {G}\,\rangle \equiv \frac{1}{2\pi}\int_{\tau_0}^{\tau_0+2\pi}
G(s, \tau)\,d\tau\equiv \overline{G}(s),\qquad\forall\ \tau_0\label{oper-1}
\end{eqnarray}

\noindent
(v)  The \emph{tilde-functions} (or purely oscillating functions) represent a special case of
$\cal{P}$-functions with zero average $\langle\widetilde G \,\rangle =0$. The \emph{bar-functions} (or
mean-functions) $\overline{G}=\overline{G}(s)$ do not depend on $\tau$. A unique decomposition
$G=\overline{G}+\widetilde{G}$ is valid.

\subsection{Asymptotic procedure}

The $\varepsilon$-dependence of $l_{ik}$ (\ref{exact-1-dless}) leads  to the presentation of matrix
$\mathbb{A}$ (\ref{matrixA}) as a series in $\varepsilon$
\begin{eqnarray}
&&\mathbb{A}=\overline{\mathbb{B}}_0+\varepsilon\delta
\widetilde{\mathbb{A}}'_0+\dots, \quad \overline{\mathbb{B}}_0\equiv{\overline{\mathbb{A}}}_0+
\delta \overline{\mathbb{A}}_{1}\label{matrix-dec}
\\
&&\overline{\mathbb{A}}_0\equiv\text{diag}\{R_1,R_2,...,R_N\},
\quad\widetilde{\mathbb{A}}'_0\equiv\begin{cases}
0  & \text{for}\ i=k,\\[-1mm]
R_{ik}\widetilde{l}_{ik}/L_{ik}^2 & \text{for}\ i\neq k
\end{cases}\nonumber
\end{eqnarray}
where we do not present the expression for $\overline{\mathbb{A}}_{1}$ since it does not affect the final
answer. In double series (with small parameters $\varepsilon$ and $\delta$) matrices
$\overline{\mathbb{A}}_0$, $\overline{\mathbb{A}}_{1}$ and $\widetilde{\mathbb{A}}'_0$ appear in the terms of
orders $\varepsilon^0\delta^0$, $\varepsilon^0\delta^1$, and $\varepsilon^1\delta^1$.

The introduction of  fast time variable $\tau$ and slow time variable $s$ represents the crucial step of our
asymptotic procedure. We choose:
\begin{eqnarray}\nonumber\label{chain0}
&& \tau=t,\qquad s=\varepsilon^2 t
\end{eqnarray}
This choice can be justified by the same distinguished limit arguments as in
\cite{VladimirovMHD}. Here we present this choice without proof, however the most important part of this
proof (that this choice  leads to a valid asymptotic procedure) is exposed and exploited below. After the use
of  the chain rule
\begin{eqnarray}\label{chain}
&& d/dt=\partial/\partial\tau+\varepsilon^2\partial/\partial s
\end{eqnarray}
we  accept (temporarily) that $\tau$ and $s$ represent two independent variables. The substitution of
(\ref{chain}),(\ref{matrix-dec}) into (\ref{exact-1-dless-matrix}) gives its two-timing form:
\begin{eqnarray}\label{exact-1-dless-matrix-ser}
&&(\overline{\mathbb{B}}_0+\varepsilon\delta
\widetilde{\mathbb{A}}'_0+\dots)(\vx_\tau+\varepsilon^2\vx_s)=\vf
\end{eqnarray}
Unknown functions are taken as regular series
\begin{eqnarray}\label{x-f-ser}
&& \vx(\tau,s)=\overline{\vx}_0(s)+\varepsilon\vx_1(\tau,s)+\dots,\quad
\vf(\tau,s)=\vf_0(\tau,s)+\varepsilon\vf_1(\tau,s)+\dots
\end{eqnarray}
where $\widetilde{\vx}_0\equiv 0$, which means that long distances of self-swimming are caused by small
oscillations: $|\overline{\vx}_0|\gg|\varepsilon\widetilde{\vx}_1(\tau,s)|$.

\subsection{Successive approximations}

The successive approximations of (\ref{exact-1-dless-matrix-ser}),(\ref{x-f-ser}) yield:

\noindent \emph{Terms of order} $\varepsilon^0=1$:\  $\vf_0\equiv 0$;

\noindent  \emph{Terms of order}  $\varepsilon^1$:  \ $\overline{\mathbb{B}}_0\vx_{0\tau}=\vf_1$;
its average gives $\overline{\vf}_1\equiv 0$ and the oscillating part is
\begin{eqnarray}\label{one-tilde}
&&\overline{\mathbb{B}}_0\widetilde{\vx}_{1\tau}=\widetilde{\vf}_1;
\end{eqnarray}

\noindent \emph{Terms of order} $\varepsilon^2$: \ $
\overline{\mathbb{B}}_0\widetilde{\vx}_{2\tau}+\delta\widetilde{\mathbb{A}}'_0\widetilde{\vx}_{1\tau}+
\overline{\mathbb{B}}_0\overline{\vx}_{0s}=\vf_2
$; its averaged part is
\begin{eqnarray}\label{two-aver}
&&\overline{\mathbb{B}}_0\overline{\vx}_{0s}+
\delta\langle\widetilde{\mathbb{A}}'_0\widetilde{\vx}_{1\tau}\rangle=\overline{\vf}_2
\end{eqnarray}
Force $\overline{\vf}_2$ can be excluded from (\ref{two-aver}),(\ref{exact-1aa}) as:
\begin{eqnarray}\label{two-aver-a}
&&\vI\cdot\overline{\mathbb{B}}_0\overline{\vx}_{0s}+
\delta\vI\cdot\langle\widetilde{\mathbb{A}}'_0\widetilde{\vx}_{1\tau}\rangle=0
\end{eqnarray}
The averaged self-propulsion motion means that the rate of changing is $\overline{x}_{0is}=\overline{X}_{0s}$
with the same function $\overline{X}_0(s)$ for all spheres; therefore we write
$\overline{\vx}_{0s}=\overline{X}_{0s}\vI$. Hence (\ref{two-aver-a}) gives
\begin{eqnarray}\label{vel-self-prop}
&&\overline{X}_{0s}=
-\delta\frac{\vI\cdot\langle\widetilde{\mathbb{A}}'_0\widetilde{\vx}_{1\tau}\rangle}{\vI\cdot\overline{\mathbb{A}}_0
\vI}
\end{eqnarray}
where in the denominator matrix $\overline{\mathbb{B}}_0$  is replaced with $\overline{\mathbb{A}}_0$, since
we consider only the main (linear  in $\delta$) term in (\ref{vel-self-prop}). Expression
(\ref{vel-self-prop}) still contains unknown functions $\widetilde{\vx}_{1\tau}$ which can be determined from
(\ref{one-tilde}) with the use of constraints (\ref{exact-1AAA}),(\ref{exact-1aa}). Indeed, equation
(\ref{one-tilde}) (with the terms required for linear in $\delta$ precision in (\ref{vel-self-prop})) gives:
\begin{eqnarray}
&&\widetilde{\vx}_{1\tau}=(\overline{\mathbb{A}}_0)^{-1}\widetilde{\vf}_1\equiv\widetilde{\vg},
\quad\widetilde{\vg}=(\widetilde{g}_1,\widetilde{g}_2, ...,
\widetilde{g}_N)\label{tilde-x}\\
&&\widetilde{g}_i\equiv\widetilde{f}_{1i}/R_i,\quad
(\overline{\mathbb{A}}_0)^{-1}=\text{diag}\{1/R_1,1/R_2,\dots,1/R_N\}\nonumber
\end{eqnarray}
One can see that (\ref{tilde-x}),(\ref{exact-1AAA}) yield
$\widetilde{x}_{\alpha+1,\tau}-\widetilde{x}_{\alpha,\tau}=\widetilde{g}_{\alpha+1}-\widetilde{g}_\alpha=
\widetilde{l}_{\alpha\tau}$,
while (\ref{exact-1aa}) leads to $\sum_i R_i\widetilde{g}_i=0$. Both restrictions can be written with the use
of $N\times N$ constraint matrix $\mathbb{C}$:
\begin{eqnarray}\nonumber\label{constraints-tilde}
&& \mathbb{C}\widetilde{\vg}=\widetilde{\vl}_\tau,\ \
 \mathbb{C}\equiv
\left(
  \begin{array}{cccccc}
    -1 & 1 & 0 &\dots &0 &0 \\
    0 & -1 & 1 &\dots &0 &0 \\
    \dots &\dots &\dots &\dots &\dots &\dots\\
    0 & 0 & 0 &\dots &-1 &1 \\
    R_1 &R_2 &R_3 &\dots &R_{N-1} &R_N\\
  \end{array}
\right),\ \
\widetilde{\vl}\equiv
\left(
\begin{array}{c}
    \widetilde{l}_{1}  \\
     \widetilde{l}_{2} \\
    \dots \\
    \widetilde{l}_{N-1} \\
    0\\
  \end{array}
\right)
\end{eqnarray}
The substitution of its inverse form
\begin{eqnarray}\label{tilde-xC}
&&\widetilde{\vx}_{1\tau}=\mathbb{C}^{-1}\widetilde{\vl}_\tau
\end{eqnarray}
into (\ref{vel-self-prop}) yields
\begin{eqnarray}\label{HER}
&&\overline{X}_{0s}= -\frac{\delta}{\rho}\vI\cdot\langle\widetilde{\mathbb{A}}'_0
\mathbb{C}^{-1}\widetilde{\vl}_{\tau}\rangle
\end{eqnarray}
where
\begin{eqnarray}\label{M-1}
&&(-1)^{N+1}\rho\mathbb{C}^{-1}\equiv
\left(
  \begin{array}{cccccc}
    \rho_1-\rho & \rho_2-\rho & \rho_3-\rho &\dots &\rho_{N-1}-\rho &1 \\
    \rho_1 & \rho_2-\rho & \rho_3-\rho &\dots &\rho_{N-1}-\rho &1 \\
    \rho_1 & \rho_2 &\rho_3-\rho &\dots &\rho_{N-1}-\rho &1\\
    \dots &\dots &\dots &\dots &\dots &\dots\\
    \rho_1 & \rho_2 & \rho_3 &\dots &\rho_{N-1}-\rho &1 \\
    \rho_1 &\rho_2 &\rho_3 &\dots &\rho_{N-1} &1\\
  \end{array}
\right)\\
&& \rho_k\equiv\sum_{i=1}^{k}R_i,\ k\geq 1;\quad \rho\equiv \rho_N,\quad
\vI\cdot\overline{\mathbb{A}}_0\vI=\rho
\nonumber
\end{eqnarray}
The expression for  inverse matrix (\ref{M-1}) can be checked by direct calculations  of  product
$\mathbb{C}^{-1}\mathbb{C}$; these calculations become particularly simple if the matrix in the right hand
side of (\ref{M-1}) is decomposed into two matrices: one containing all $\rho$'s, and another with identical
rows $(\rho_1,\rho_2,\rho_3,\dots, \rho_{N-1}, 1)$. It is worth to emphasise that the presented analytic
procedure is especially simple, since in order to calculate  the self-propulsion velocity
(\ref{vel-self-prop}) one needs only to know the main approximation $\widetilde{\vx}_{1}$  for mutual
oscillations of spheres, while this approximation is completely described by simple equations
(\ref{tilde-x}),(\ref{tilde-xC}).

\subsection{Self-propulsion velocity}

One can see from (\ref{HER}) that $\overline{X}_{0s}=\vI\cdot\overline{\vx}_s/N= O(\delta)$. Hence the order
of magnitude of dimensionless physical velocity is:
$$
\overline{V}_0\equiv\vI\cdot\overline{\vx}_t/N=\varepsilon^2
\overline{X}_{0s}=O(\varepsilon^2\delta)
$$
The substitution  of $\widetilde{\mathbb{A}}'_0$ (\ref{matrix-dec}) and $\mathbb{C}^{-1}$ (\ref{M-1}) into
(\ref{HER}) and subsequent algebraic transformations lead to
\begin{eqnarray}
&&\overline{V}_0 = \frac{\varepsilon^2\delta}{\rho^2}\sum_{i<k<l} \overline{G}_{ikl}\label{vel-self-prop-final+}\\
&&\overline{G}_{ikl}\equiv 2
R_iR_kR_l\left(\frac{1}{L_{ik}^2}+\frac{1}{L_{kl}^2}-\frac{1}{L_{il}^2}\right)\langle
\widetilde{l}_{ik}\widetilde{l}_{kl\tau}\rangle\label{G}
\end{eqnarray}
where the sum (\ref{vel-self-prop-final+}) is taken over all possible triplets $(i,k,l):\ 1\leq i<k<l\leq N$.
In (\ref{G}) one can also take $2\langle
\widetilde{l}_{ik}\widetilde{l}_{kl\tau}\rangle=\langle
\widetilde{l}_{ik}\widetilde{l}_{kl\tau}-\widetilde{l}_{ik\tau}\widetilde{l}_{kl}\rangle$, which can be
proved by integration by parts. The formulae (\ref{vel-self-prop-final+}), (\ref{G}) have been obtained for
$N=3,4,5$ by explicit analytical calculations  and for any $N$ by the method of mathematical induction. These
calculations are straightforward but rather cumbersome to be presented here. However, as soon as
(\ref{vel-self-prop-final+}) and (\ref{G}) are known, they can be verified by separate calculations of all
terms proportional to ${1}/{L_{ik}^2}$ for each particular pair $i,k$. For example, for $i=1, k=2$ the
relevant part of matrix $\widetilde{\mathbb{A}}'_0\equiv
\widetilde{A}'_{ik}$ (\ref{matrix-dec}) contains only  $\widetilde{A}'_{12}=\widetilde{A}'_{21}=
 R_{12}\widetilde{l}_{1}/L_{12}^2$, while all other components are zero. The related part of (\ref{HER}) can
be easily calculated; it leads to the same expression as the corresponding extraction from
(\ref{vel-self-prop-final+}),(\ref{G}), which in this case contains only $(N-2)$ terms with $l=3,4,\dots, N$.

The number of terms/triplets in (\ref{vel-self-prop-final+}) rapidly increases with $N$: for a three-swimmer
the sum (\ref{vel-self-prop-final+}) contains the only triplet, for a four-swimmer -- four triplets, for a
five-swimmer -- 10 triplets, while for a ten-swimmer the number of triplets grows up to 120. In general
(\ref{vel-self-prop-final+}) contains $N!/[(N-3)!3!]$ triplets. It is important to emphasise that
(\ref{vel-self-prop-final+}) contains \emph{all triplets} in a micro-robot, not only triplets of the
neighbouring spheres. If one takes into account only the triplets of the neighbouring spheres (say, (1,2,3)
and (2,3,4) out of $(1,2,3), (1,2,4), (1,3,4),(2,3,4)$ for $N=4$), then they give only the rough estimation
of $\overline{V}_0$, which can be misleading since different correlations $\langle
\widetilde{l}_\alpha\widetilde{l}_{\beta\tau}\rangle$ can have different values (see below).

Function $\overline{G}_{123}$ (\ref{G}) has been introduced  by \cite{Golestanian} and studied by
\cite{Yeomans, Lefebvre, Golestanian1} in the context of a three-sphere micro-robot. We call
$\overline{G}_{ikl}$ \emph{a Golestanian function}. Formulae
(\ref{HER}),(\ref{vel-self-prop-final+}),(\ref{G}) give the main result of this paper: the self-swimming
velocity of an $N$-sphere micro-robot represents a linear combination of  Golestanian functions for all
available triplets.

\subsection{Examples of homogeneous micro-robots: power, velocity, and efficiency }

The explicit formulae (\ref{vel-self-prop-final+}), (\ref{G}) can be used to obtain  physically interesting
results (optimal strokes,  required power, related forces, efficiency,
\emph{etc.}) for various $N$-sphere swimmers.
We briefly address some of these questions below. In all these examples, we consider only homogeneous
micro-robots, consisting of equal spheres $R_i=1$ and of equal upstretched arms $L_\alpha=1$.

The scalar product  of the main equation (\ref{exact-1-dless}) and $\vx_t$ leads to the average power of a
micro-robot
$$\overline{\mathcal{P}}\equiv\langle \vf\cdot\vx_{t} \rangle=\varepsilon^2 \langle
\widetilde{\vx}_{1\tau}^2\rangle +O(\varepsilon^2\delta)
$$
where we have taken into account that $\vx_t=\varepsilon\widetilde{\vx}_{1\tau}+O(\varepsilon^2)$, which
follows from (\ref{x-f-ser}) and (\ref{chain}). Another expression $\overline{\mathcal{P}}_s=\rho
\overline{V}_0^2$  represents the power, which is required to drag  a micro-robot
with velocity $\overline{V}_0$ in the absence of its oscillations (when the main approximation for the
dimensionless Stokes's friction force is $-\rho \overline{V}_0$).  Lighthill's swimming efficiency (see
\cite{Koelher}) is the ratio $\mathcal{E}\equiv{\overline{\mathcal{P}}_s}/{\overline{\mathcal{P}}}$, which in
our case is
\begin{eqnarray}
&&\mathcal{E}\simeq\frac{\varepsilon^2\delta^2}{\rho^3}
\frac{\left(\sum_{i<k<l} \overline{G}_{ikl}\right)^2}{\langle
\widetilde{\vx}_{1\tau}^2\rangle}
\label{Eff}
\end{eqnarray}
where $\widetilde{\vx}_{1\tau}$ is determined by (\ref{tilde-xC}).

 For \textbf{\emph{a three-swimmer}}, eqns.(\ref{vel-self-prop-final+}),(\ref{G}) yield:
\begin{eqnarray}
&&\overline{V}_0=\frac{\varepsilon^2\delta}{9}\overline{G}_{123},\quad
\overline{G}_{123}=\frac{7}{2}\langle
\widetilde{l}_1 \widetilde{l}_{2\tau}\rangle\label{3-vel}
\end{eqnarray}
Further simplification can be achieved if we accept that oscillations of both arms are harmonic and have
equal amplitudes
\begin{eqnarray} l_\alpha=\cos(\tau+\varphi_\alpha),\quad\text{then}\quad 2\langle\widetilde{l}_1
\widetilde{l}_{2\tau}\rangle=\sin(\varphi_1-\varphi_2)\label{3-harm}
\end{eqnarray}
with constant phases $0\leq\varphi_\alpha\leq 2\pi$. Substitution of (\ref{3-harm}) into (\ref{3-vel}) and
(\ref{Eff}) gives
\begin{eqnarray}
&&\overline{V}_0=\varepsilon^2\delta\frac{7}{36}\sin\phi,\quad
\mathcal{E}=\left(\frac{7\varepsilon\delta}{12}\right)^2\frac{\sin^2\phi}{2+\cos\phi},\quad
\phi\equiv\varphi_1-\varphi_1\nonumber\label{3-vel-phase}
\end{eqnarray}
which shows that their  maxima take place at different $\phi$:
\begin{eqnarray}
&&\max\overline{V}_0\simeq 0.19\varepsilon^2\delta \ \text{at}\ \phi=\pi/2\simeq
1.57\label{3-maxima}\\
&&\max \mathcal{E}=0.182\varepsilon^2\delta^2\ \text{at}\ \phi=1.80\nonumber
\end{eqnarray}
Similar consideration for \textbf{\emph{a four-swimmer}}  yields
\begin{eqnarray}
&&\overline{V}_0=\frac{\varepsilon^2\delta}{16}(\overline{G}_{123}+\overline{G}_{124}+\overline{G}_{134}
+\overline{G}_{234})
\nonumber\label{vel-self-prop-final++}
\end{eqnarray}
which (with the use of (\ref{3-harm})) leads to
\begin{eqnarray}
&&\overline{V}_0=\varepsilon^2\delta\frac{7}{64} \mathcal{{S}}_4(\phi,\psi),\quad
\mathcal{{S}}_4(\phi,\psi)\equiv(1+C)(\sin
\phi+\sin\psi)+2C\sin(\phi+\psi)\nonumber
\label{V04}
\end{eqnarray}
where $C=41/63\simeq0.65$, $\phi\equiv\varphi_1-\varphi_2$, and $\psi\equiv \varphi_2-\varphi_3$. Then
(\ref{Eff}) can be expressed as:
\begin{eqnarray}
&&\mathcal{E}=\left(\frac{7\varepsilon\delta}{16}\right)^2\frac{\mathcal{S}_4^2(\phi,\psi)}
{5+2\cos\phi+2\cos\psi+\cos(\phi+\psi)}\nonumber\label{4-eff}
\end{eqnarray}
The computations show that:
\begin{eqnarray}\label{4-maxima}
&&\max\overline{V}_0\simeq0.44\varepsilon^2\delta\ \text{at}\ \phi\simeq\psi\simeq 1.10\\
&&\max \mathcal{E}\simeq 0.55\varepsilon^2\delta^2\ \text{at}\ \phi\simeq\psi\simeq 1.38\nonumber
\end{eqnarray}

For \textbf{\emph{a five-swimmer}}, one can write
\begin{eqnarray}
&&\overline{V}_0=\frac{\varepsilon^2\delta}{25}(\overline{G}_{123}+\overline{G}_{124}+\overline{G}_{125}+
\overline{G}_{134}+\overline{G}_{135}+\overline{G}_{145}+\overline{G}_{234}+
\overline{G}_{235}+\overline{G}_{245}+\overline{G}_{345})
\nonumber\label{vel-5}
\end{eqnarray}
which leads to:
\begin{eqnarray}
&&\overline{V}_0=\frac{\varepsilon^2\delta}{50}\mathcal{S}_5,\quad   \mathcal{S}_5\equiv
(4a+b+c)(\sin\phi+\sin\chi)+(5a+2b)\sin\psi+\nonumber
\label{V05}\\
&& (a+2b+c)[\sin(\phi+\psi)+\sin(\psi+\chi)]+(a+2c)\sin(\phi+\psi+\chi);\nonumber\\
&&a=7/8,\ b=41/18,\ c=151/72;\quad\phi\equiv\varphi_1-\varphi_2,\ \psi\equiv \varphi_2-\varphi_3,
\ \chi\equiv \varphi_3-\varphi_4\nonumber
\end{eqnarray}
Then (\ref{Eff}) takes form:
\begin{eqnarray}
&&\mathcal{E}=\left(\frac{\varepsilon\delta}{10}\right)^2\frac{\mathcal{S}_5^2}{\mathcal{P}_5},\quad
\mathcal{P}_5\equiv10+3\cos\phi+4\cos\psi+3\cos\chi+
\label{5-eff}\\
&&2\cos(\phi+\psi)+2\cos(\psi+\chi)+\cos(\phi+\psi+\chi)
\nonumber
\end{eqnarray}
The computations give:
\begin{eqnarray}\label{5-maxima}
&&\max\overline{V}_0\simeq0.77\varepsilon^2\delta\ \text{at}\ \phi\simeq\chi\simeq 0.86,\ \psi\simeq 0.83\\
&&\max \mathcal{E}\simeq 1.00\varepsilon^2\delta^2\ \text{at}\ \phi\simeq\chi\simeq 1.14,\ \psi\simeq
1.08\nonumber
\end{eqnarray}
The numerical results (\ref{3-maxima})-(\ref{5-maxima}) show that both $\max\overline{V}_0$ and $\max
\mathcal{E}$ grow when $N$ increases. For reasonably small values of parameters (say, $\varepsilon\simeq 0.2$ and $\delta\simeq 0.2$) we have $\max
\mathcal{E}\sim 0.1\%$, hence the efficiency of considered micro-robots is low.

In order to compare the velocities of micro-robots and micro-organisms we use the dimensional variables, in
which $\max\overline{V}_0^*\sim\omega^* L^*
\varepsilon^2\delta$; it  shows that (for typical stroke frequency of self-swimming microorganisms, which is about
several $Hz$, see \cite{PedKes, Pedley, Polin, VladPedl}) a micro-robot can move itself with the rate  about
$10\%$ percents of its own size per second. This estimation is $20\div 40$ times lower than a similar value
for natural micro-swimmers, see
\cite{VladPedl}; it  shows again low efficiency of considered micro-robots.
If we suggest that the function $\max\overline{V}_0(N)$ grows with a similar rate (as it has been calculated
for $N=3,4,5$), then micro-robots with $N=8\div 10$ could swim with a speed, similar to that of
micro-organisms.

\section{Discussion}

1. Our approach (based on the two-timing method and a distinguished limit) is technically different from all
previous methods employed in the studies of micro-robots. The possibility to derive explicit formulae for an
$N$-sphere micro-robot shows its strength and analytical simplicity. The used version of the two-timing
method has been developed in \cite{Vladimirov0,Vladimirov1,VladimirovMHD}.

2. One can see that $\overline{V}_0=O(\varepsilon^2\delta)$ (\ref{vel-self-prop-final+}), which is the same
as the result by \cite{Golestanian, Golestanian1} for a three-sphere swimmer. At the same time our choice of
slow time $s=\varepsilon^2 t$ (\ref{chain}) agrees with classical studies of self-propulsion for low Reynolds
numbers, see \cite{Taylor, Blake, Childress}, as well as geometric studies of \cite{Wilczek}.

3. The `triplet' structure of a  formula for self-propulsion velocity (\ref{vel-self-prop-final+})  can be
expected without any calculations, on the base of the result for $N=3$. Indeed, if we are interested in the
main term of the order $\varepsilon^2\delta$, then only triple interactions can be taken into account, as
they have been described by \cite{Golestanian}. For example, the interactions between four spheres produce a
term of the next order $O(\varepsilon^3\delta)$ in  $\overline{V}_0$, where the multiplier $\varepsilon^3$
appears from the motion of three  arms involved.

4. In our examples, all arms move harmonically (\ref{3-harm}); it does not provide the maximum of
$\overline{V}_0$. For example, for a three-sphere robot $\overline{V}_0\sim
\langle\widetilde{l}_1\widetilde{l}_{2\tau}\rangle$ (\ref{3-vel}). Since $\widetilde{l}_1$ and
$\widetilde{l}_{2\tau}$ represent mutually independent $2\pi$-periodic in $\tau$ functions,  it is clear that
the maximum of this correlation appears when these functions coincide or proportional to each other. If
$\max\langle\widetilde{l}_1\widetilde{l}_{2\tau}\rangle$ is calculated under the constraint of fixed
amplitudes (which is natural for realistic experimental devices  of variable arm lengths), then one can find
that the theoretical maximum of this correlation  is $2/\pi$, which is higher than $1/2$ for  harmonic
oscillations (\ref{3-harm}). Such an improvement will increase with the growth of $N$. In particular,
non-harmonic periodic $\widetilde{l}_{\alpha}(\tau)$, providing the optimal strokes, have been discovered in
computational studies of  four-sphere  micro-robots by \cite{Yeomans}.

5. In our study we build an asymptotic procedure with two small parameters: $\varepsilon\to 0$ and $\delta\to
0$. Such a setting usually requires the consideration of different asymptotic paths on the plane
$(\varepsilon,\delta)$ when, say $\delta=\delta(\varepsilon)$. In our case we can avoid such  consideration,
since  small parameters appear (in the main order) as a product $\varepsilon^2\delta$.

6. The mathematical justification of the presented results by the estimation of an error in the original
equation can be performed similar to \cite{VladimirovX1,VladimirovX2}. One can also derive the higher
approximations of $\overline{V}_0$, as it has been done by
\cite{VladimirovX1,VladimirovX2} for different cases. The higher approximations can be useful for the studies
of motion with $\overline{V}_0\equiv 0$ (say, if all correlations involved in (\ref{vel-self-prop-final+})
are zero).

7. In the literature quoted in \emph{Introduction} one can find interesting discussions about the physical
mechanism of self-propulsion of micro-robots. A  clear illustration of this mechanism is given by
\cite{Avron}. At the same time one can  notice that the self-propulsion of deformable bodies
in inviscid fluid represents a classical topic, see \emph{e.g.} \cite{Saffman, Newton}. It is interesting to
note that the qualitative explanations of self-propulsion in an inviscid fluid and in self-propulsion in
creeping flow can be seen as the same if one replaces the term
\emph{virtual mass} (for an inviscid fluid) to \emph{viscous drag} (for creeping flows).


\begin{acknowledgments}
The author is grateful to Profs. A.D.Gilbert, R.Golestanian,  K.I. Ilin, H.K.Moffatt, and J. Pitchford for
useful discussions.
\end{acknowledgments}

\end{document}